\documentclass[a4paper, 10pt, conference]{IEEEtran}      % Use this line for a4 paper

\IEEEoverridecommandlockouts
\usepackage[utf8]{inputenc} 
\usepackage[T1]{fontenc}    
\usepackage{hyperref}       
\usepackage{url}          
\usepackage{booktabs}      
\usepackage{amsfonts}       
\usepackage{nicefrac}       
\usepackage{microtype}     
\usepackage{graphics} 
\usepackage{graphicx}
\usepackage{dblfloatfix} 
\usepackage{caption}
\usepackage{subcaption}
\usepackage{amsmath}
\usepackage{amssymb}
\usepackage{catchfile}
\usepackage{totcount}
\usepackage[ruled,vlined]{algorithm2e}
\usepackage{enumitem}
\usepackage{multirow}
\begin{document}

\title{\normalsize {\bf
Data Enforced: An Exploratory Impact Analysis of Automated Speed Enforcement in the District of Columbia}\\
{\small *accepted for publication at the 24th IEEE Intelligent Transportation Systems Conference (ITSC 2021) 
\thanks{\copyright 2021 IEEE Personal use of this material is permitted. Permission from IEEE must be obtained for all other uses, in any current or future media, including reprinting/republishing this material for advertising or promotional purposes, creating new collective works, for resale or redistribution to servers or lists, or reuse of any copyrighted component of this work in other works.}}
}

\author{

\IEEEauthorblockN{\bf Awad Abdelhalim}
\IEEEauthorblockA{\textit{Department of Civil and Environmental Engineering} \\
{\small \textit{Virginia Polytechnic Institute and State University}} \\
{\small Blacksburg, Virginia, USA} \\
{\small atarig@vt.edu}}

\and

\IEEEauthorblockN{\bf Linda Bailey\textsuperscript{1}, Emily Dalphy\textsuperscript{1}, and Kelli Raboy\textsuperscript{2}}
\IEEEauthorblockA{\textit{\textsuperscript{1} Vision Zero Initiative, \textsuperscript{2} ITS Program Manager} \\
{\small \textit{District Department of Transportation}} \\
{\small Washington, District of Columbia, USA} \\
{\small linda.bailey, emily.dalphy, kelli.raboy <@dc.gov>}}
}
\maketitle

%%%%%%%%%%%%%%%%%%%%%%%%%%%%%%%%%%%%%%%%%%%%%%%%%%%%%%%%%%%%%%%%%%%%%%%%%%%%%%%%
\begin{abstract}
In 2015, the District of Columbia framed a Vision Zero mission and action plan, with a target of achieving zero traffic fatalities by 2024. This study examines the impacts of Automated Speed Enforcement (ASE) and its role in achieving the goals of Vision Zero. Independent datasets containing detailed information about traffic crashes, ASE camera locations and citation records, and driving speeds across the District's streets were collected, combined, and analyzed to identify patterns and trends in crashes, speed limit violations, and speeding behavior before and after the ASE camera installation. The results of this exploratory analysis confirm the safety benefits of ASE systems in Washington, D.C. The study also provides a blueprint for the different means of evaluating the short-term impact of ASE systems using different data sources which can aid practitioners in better evaluating existing systems and support the decision making process regarding future installations.
\end{abstract}

%\begin{keywords}
%Automated speed enforcement, 
%\end{keywords}

%%%%%%%%%%%%%%%%%%%%%%%%%%%%%%%%%%%%%%%%%%%%%%%%%%%%%%%%%%%%%%%%%%%%%%%%%%%%%%%%
\section{Introduction}

Traffic crashes are one of the leading causes of death in the World. In the United States, about 40,000 lives are lost annually due to car crashes. In the District of Columbia, over 7,000 people are involved in injury crashes every year; of those crashes, over 300 result in serious, life-altering injuries, and in 2019, 27 have resulted in death. As the District works to create safe streets for everyone, speed management is key. Lower speeds allow drivers more time to avoid crashes, and mitigate the severity of any crashes that do occur. Automated speed enforcement (ASE) is a critical part of reducing speeds, particularly on large arterial roadways in the off-peak time period, when traffic is lighter.

\subsection{Vision Zero}
The District of Columbia launched Vision Zero in 2015 to inspire and transform DC’s roadway safety efforts, and set sights for a goal of zero fatalities or serious injuries on the District's roads. Since then, DDOT has shifted to designing streets that are safe for everyone, however they choose to move around the city. The Vision Zero Division has developed and implemented a myriad of safety projects that address the concerns of all the different groups of roadway users. Working closely with community members to identify problems, as well as using data on past crashes, Vision Zero focuses on managing speeds and conflict points, and promoting safer modes of travel, such as transit, walking and cycling. However, challenges remain in our road safety program. Vulnerable road users like pedestrians and cyclists remain over-represented in the District's traffic fatalities and severe injuries, and some of the most at-risk populations for other types of violence are also suffering disproportionately from traffic deaths and injuries. ASE is a key pathway to achieving lower speeds and safer streets for the District. ASE is often publicly criticized as a revenue tool for government agencies rather than a safety measure. This study aims to assess the District's ASE system both in terms of it's impact on traffic safety as well as the rate and trends in issued citations.

\subsection{Related Work}
Various studies have been carried out through the years on Automated Traffic Enforcement systems, mainly red light cameras (RLCs) and ASE systems. There has been a consensus in the conclusions that ATEs lead to a reduction in the overall number of crashes, and a significant reduction in the severity of crashes. Some studies have concluded that RLCs may be associated with an increase in rear-end crashes as the drivers come to abrupt stops in avoidance of citations. Other studies concluded that ASE may lead to a local reduction in speeds, but lead to drivers accelerating downstream of the enforcement site to make up the time lost driving at lower speeds.

The vast majority of studies on ASE conclude that the overall benefits overshadow the cons. Researchers have found ASE cameras to lead to significant reductions in the frequency and severity of crashes in their vicinity, notably \cite{wilson} , \cite{hauer} \cite{depauw}. A study by Shin et al. \cite{shin} concluded that ASE cameras lead to great reduction in all types of crashes but rear-end crashes. Cunningham et al. \cite{cunningham} concluded that ASEs not only lead reduced frequency of crashes, but also positively affect the overall speeding behavior. Some studies have concluded that the effect of the ASE cameras decays over time \cite{carnis}. Other researchers also propose the use of Automated Section Speed Enforcement monitoring average speed over a closed section as opposed to a fixed camera location and have concluded that it produces substantial benefits \cite{montella}, \cite{soole}. A previous study evaluating automated speed enforcement in the District of Columbia \cite{jonathan} has concluded that the system produces substantial reduction in the District's car crash frequency and in the severity of crashes.

\begin{table*}[!t]
  \caption{Data Sources Used In The Analysis}
  \label{tab:sources}
  \centering
  \renewcommand{\arraystretch}{1.3}
   \begin{tabular}{lccc} \hline
        \multicolumn{1}{c}{} & \multicolumn{3}{c}{\textbf{Details}} \\ \cline{2-4}
        \textbf{Dataset} & \textbf{Provider} & \textbf{Information Contained} & \textbf{Number of Records} \\ \hline
        \textbf{DC Crashes} & PSI & Traffic crashes details & 92,360 \\ \hline
        \textbf{Camera Locations} & MPD & Geolocations of ASE cameras & 84 \\ \hline
        \textbf{Camera Citations} & MPD & Hourly violations and citations records & 72,965 \\ \hline
        \textbf{Probe Vehicle} & INRIX & Travel time and speed & 1,997,410 \\ \hline
    \end{tabular}
\end{table*}

\begin{figure*}[!b]
\begin{subfigure}{.5\textwidth}
  \includegraphics[width= 3.4in, height= 2.4in]{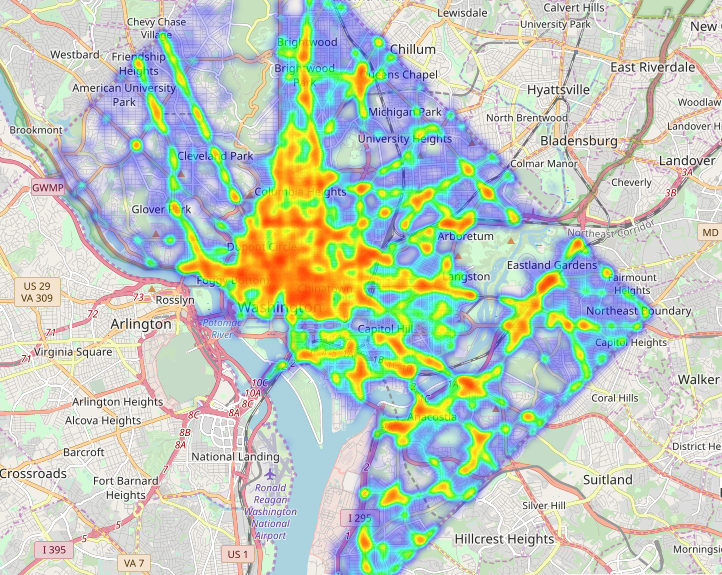}
  \caption{\label{fig:sub1} All car crashes in DC.}
\end{subfigure}
\begin{subfigure}{.5\textwidth}
  \includegraphics[width= 3.4in, height= 2.4in]{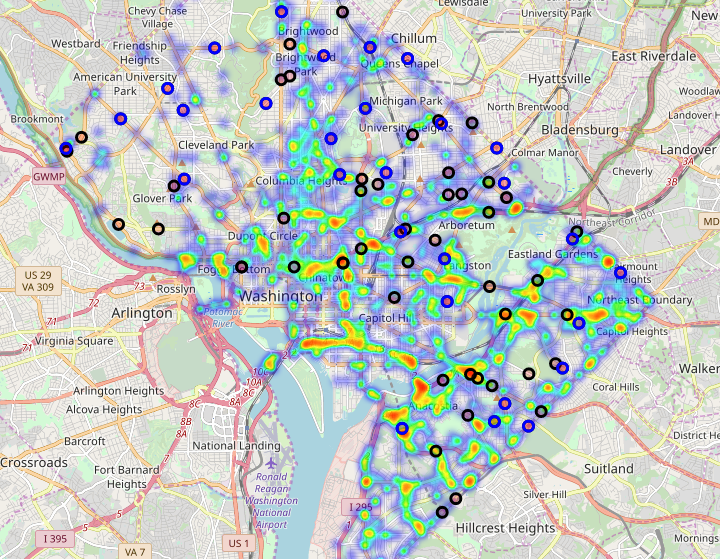}
  \caption{ \label{fig:sub2} Crashes due to speeding.}
\end{subfigure}
\caption{\label{fig:heatmaps} Heatmaps of traffic crashes in DC.}
\end{figure*}

\subsection{Objective}
The objective of this study is to quantify the impacts of ASE on crash frequencies, severity, and speeding behavior in select locations within the District of Columbia, aiming to answer the following questions: (1) is the ASE system beneficial to the overall traffic safety in the District, specifically in terms of reducing crash frequencies and severity? (2) what percentage of roadway users in the District are receiving citations due to ASE? And (3) does the ASE system help in changing overall speeding behavior along corridors or is the speed reduction only occurring in the site of the enforcement cameras?

\section{Methodology}
\vspace{5pt}

\begin{figure*}[!t]
    \centering
    \includegraphics[width= \textwidth, height= 2.7in]{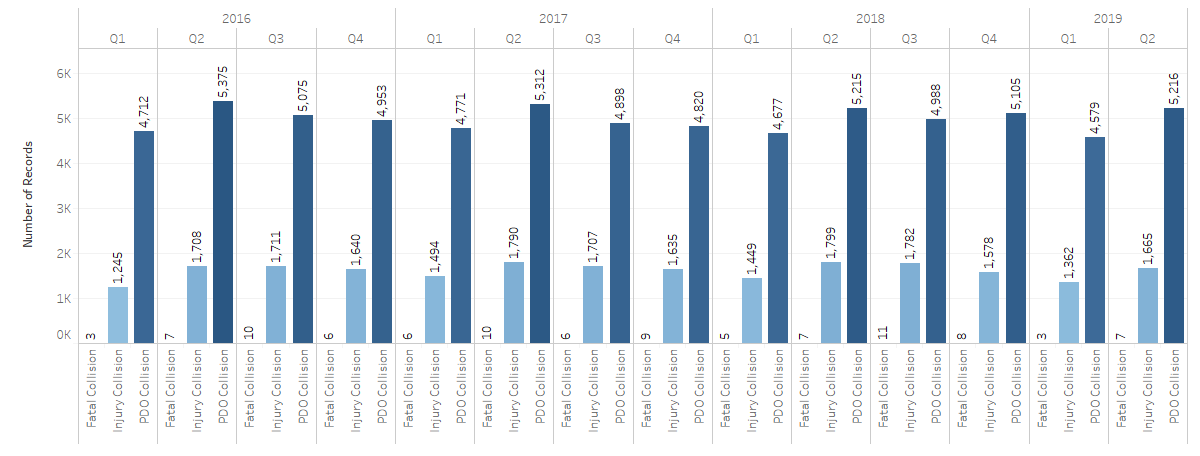}
    \caption{\label{fig:counts} Quarterly crash counts and severity of collisions during the period of analysis (2016 to mid-2019).}
\end{figure*}

\subsection{Data Sources}
To conduct the aforementioned analyses, we gathered, explored and analyzed multiple, independent, datasets collected by the District Department of Transportation (DDOT) and the Metropolitan Police Department (MPD) that include detailed data on traffic crashes, ASE camera locations and citations. We also used a probe vehicle dataset collected by a third party containing the travel speeds across the District's streets. The datasets used in this study and the sources they were obtained from are listed in table \ref{tab:sources}. Crash data is available publicly through the OpenDC data portal based on police reports, but DDOT contracts a third party vendor, Precision Systems, Inc (PSI) to further verify and clean the data. Hence, the crash data was used from PSI's database. The crash dataset includes detailed information on the crash time and date, geo-location, severity, vehicles involved, influencing factors, alongside the number and types of injuries. 

Data quality issues came into question as the District's traffic crashes databases have experienced severe under-reporting prior to changes in the reporting system made in August of 2015 which have significantly enhanced the quantity and quality of data collected thereafter. In order to avoid data bias due to under-reported crashes prior to 2016, the crash data used in this analysis included only crashes that took place from the beginning of 2016 until June of 2019. This has had a significant impact on the number of study locations viable for a before/after analysis, as the majority of the District's 84 speed cameras have been installed prior to that. For the purpose of this analysis, 29 locations have been selected where camera installation or enforced speed limit change has taken place between January of 2017 and mid-2018. This has limited the period of comparative analysis for this study to a single year before and after camera installation, which, albeit sub-optimal, could provide the high-level insight on the direct impact of ASE sought after by this exploratory study and ensures the crash data does not contain the aforementioned biases. Figure \ref{fig:heatmaps} shows the geo-spatial distribution of crashes in the District. Figure \ref{fig:heatmaps} (b) shows the crashes reported as resulting from speeding, with the circles denoting the location of ASE cameras. The blue circles denote the cameras included in the before/after analysis of this study.

Further data was obtained from MPD that includes hourly records of passing data for all speed camera locations between 2013 to mid-2019. The dataset includes the hourly counts of violators and non-violators, and their respective average passing and violating speeds. This dataset allowed the analysis of the trend in citations issued by the ASE cameras. Furthermore, probe vehicle data that includes one-minute average speeds and travel times on predefined roadway segments in the District was obtained from INRIX, this was used for the analysis trends and patterns in speeds along roadway segments prior to and post ASE camera installation in order to quantify and visualize the impact (if any) that those cameras have on drivers behavior within those roadway segments.

\subsection{Data Pre-Processing and Exploration}

All dataset pre-processing, cleaning, and aggregation was carried out in SQL. During the period of interest between January of 2016 - June 2019, a total of 92,360 crashes have taken place within the District of Columbia. Figure \ref{fig:counts} shows the quarterly breakdown and severity of the crashes.

The geo-spatial mapping of crashes and speed camera locations shown in figure \ref{fig:heatmaps} indicated that the majority of crashes that are reported as speed-related occur in locations with no speed camera installed. The speed camera location are indicated by circles, where the blue circles indicate the camera locations of interest in the study. It is worth noting that only 2\% of crashes ($\sim$ 1,800 crashes) were reported as crashes where speeding was the causing factor. Such attribute is most likely under-reported, since the police officers who produce the crash reports arrive on-site after the crash has taken place, and would most likely report the crash as a result of speeding only if sufficient evidence exist (in terms of damage). Due to the speeds of the majority of streets in the District being at or below 30 miles per hour, this is typically under-reported. Furthermore, crashes resulting from Driving Under Influence (DUI) are not reported as speeding crashes, even if speeding was involved. Figure \ref{fig:counts} shows that there's a very small number of fatal crashes occurring in the District. Hence, the before and after analyses of crashes was conducted on all crashes, and then broken down to look specifically at injury and property damage only (PDO) crashes.

\subsection{Geo-spatial Clustering of Near-Camera Crashes}
In order to identify the crashes taking place in the vicinity of cameras, the distance from the location of each crash to the location of the 29 selected locations of interest was calculated, using haversine distance shown in equation \ref{eq:haversine}.

\begin{equation}
    \label{eq:haversine}
    d = 2r*arcsin \scriptscriptstyle{ \Bigg(\sqrt{sin^2\Big(\frac{\phi_2-\phi_1}{2}\Big)+cos(\phi_1)cos(\phi_2)sin^2\Big(\frac{\lambda_2-\lambda_1}{2}\Big)}\Bigg)}
\end{equation}

\noindent Where $\phi_1, \phi_2, \lambda_1, \lambda_2$ are respectively the latitude and longitude coordinates for the given car crash and the coordinates of ASE cameras locations.

Crashes were then deemed to belong within the vicinity of a specific speed camera location if they fall within the area of influence associated with that camera. In order to determine the area of influence for the speed cameras, a maximum distance of 750 feet was assumed. For locations where the distance to the nearest adjacent intersection was greater than 750 feet, this maximum distance was used. In locations with adjacent traffic signals, the distance was decreased in order reduce the upstream influence of traffic signals operations on travel speeds. For example, the camera located on the 3100 block of Alabama Ave NE (facing northeast-bound travel) is influenced by the upstream signal at the intersection of 32\textsuperscript{nd} Street \& Alabama Ave SE. Thus, the area of influence was reduced to 400 feet.

\begin{equation}
  \small
  \text{Camera assignment}=
  \begin{cases}
    \text{argmin}(d), & \text{if} \hspace{3pt} \text{min}(d) \leq range \\
    \text{None}, & \text{otherwise}
  \end{cases}
  \label{3}
\end{equation}

Since haversine distance is blind to topography and street layout, crashes deemed to fall in the vicinity of each camera location were further queried using street identifiers (an attribute in the crashes dataset) and the travel directions of vehicles involved in a crash, such that crashes are only assigned to an ASE camera location should they fall within the specified range from the speed camera, and at least one of the vehicles involved is traveling on the same street and in the same direction that the camera is enforcing the speed limit. This was to ensure that crashes that may have minor location accuracy errors and have occurred on crossing or nearby streets are not mistakenly taken into account. As previously mentioned, the ASE locations for the before and after crash assessment were selected on the basis of data availability from 2016 onward. Hence, all locations have at least one year of crash data before and after installation.

\section{Results}
We present the results of our analyses in terms of three different key points of interest:
\begin{enumerate}[label=(\Alph*)]
    \item Before and After Crash Counts
    \item Speed Limit Violations and Citation Rates
    \item Changes in Speeding Behavior
\end{enumerate}

\subsection{Before and After Crash Counts Comparison at Speed Camera Locations}

The analysis of the before and after crash statistics at the selected camera locations shows a downtrend in all crash types. For all the 29 locations, the average year-to-year reduction in crashes in the vicinity of the cameras was 9.35\%, 13.16\%, and 30.30\% respectively for all crashes, property damage only (PDO), and injury crashes as shown in Table \ref{tab:befaft}. Table \ref{tab:stats} shows the descriptive statistics of the before and after crash counts tallied for each of the 29 ASE camera locations. Significant reductions can be seen in average and maximum crashes near ASE camera sites after installation. The reduction in standard deviation of crash frequencies after installation also indicates that the reduction of crashes occurs across all sites.

\begin{table}[!h]
  \caption{Before/After Crash Analysis}
  \label{tab:befaft}
  \centering
  \renewcommand{\arraystretch}{1.5}
   \begin{tabular}{lccc} \hline
        \multicolumn{1}{c}{} & \multicolumn{3}{c}{\textbf{Aggregate Totals}} \\ \cline{2-4}
        \textbf{Crash Type} & \textbf{Before} & \textbf{After} & \textbf{\% Change} \\ \hline
        \textbf{All} & 107 & 97 & -9.35\% \\ \hline
        \textbf{PDO Crashes} & 76 & 66 & -13.16\% \\ \hline
        \textbf{Injury Crashes} & 33 & 23 & -30.30\% \\ \hline
    \end{tabular}
\end{table}

\begin{table}[htbp]
  \centering
  \renewcommand{\arraystretch}{2}
  \caption{Descriptive Statistics For The Total Before and After Crash Counts For The 29 ASE Camera Locations}
    \begin{tabular}{lcccccc}
    \toprule
    \multirow{2}{*}{} & \multicolumn{2}{c}{\textbf{All Crashes}} & \multicolumn{2}{c}{\textbf{PDO Crashes}} & \multicolumn{2}{c}{\textbf{Injury Crashes}} \\
\cline{2-7}          & \textbf{Before} & \textbf{After} & \textbf{Before} & \textbf{After} & \textbf{Before} & \textbf{After} \\
    \hline
    \textbf{min} & 0.00  & 0.00  & 0.00  & 0.00  & 0.00  & 0.00 \\
    \hline
    \textbf{25\%} & 1.00  & 0.00  & 0.25  & 0.00  & 0.00  & 0.00 \\
    \hline
    \textbf{50\%} & 2.00  & 2.00  & 2.00  & 1.00  & 1.00  & 0.00 \\
    \hline
    \textbf{75\%} & 4.00  & 5.00  & 3.75  & 3.00  & 2.00  & 2.00 \\
    \hline
    \textbf{90\%} & 9.60  & 7.40  & 7.50  & 6.00  & 4.20  & 2.00 \\
    \hline
    \textbf{max} & 21.00 & 12.00 & 17.00 & 9.00  & 5.00  & 4.00 \\
    \hline
    \textbf{mean} & 3.74  & 3.10  & 2.92  & 2.15  & 1.36  & 0.96 \\
    \hline
    \textbf{std} & 4.94  & 3.32  & 3.93  & 2.63  & 1.60  & 1.20 \\
    \bottomrule
    \end{tabular}%
  \label{tab:stats}%
\end{table}%

Comparing the results above to the overall year-to-year changes throughout the District showcases the positive safety impact of the automated speed enforcement system. The year-to-year changes across the different crash types doesn't follow any trend and seems almost stagnant, aside from a 5.5\% increase in injury crashes between 2016 and 2017, shown in Table \ref{tab:district}. It is worth noting that those annual statistics do not take into account any changes in traffic volume.

\begin{table}[!h]
  \caption{District-wise Year-to-Year Changes}
  \label{tab:district}
  \centering
  \renewcommand{\arraystretch}{1.75}
   \begin{tabular}{lcc} \hline
        \textbf{Crash Type} & \textbf{2016-2017} & \textbf{2017-2018} \\ \hline
        \textbf{All} & 0.05\% & 0.67\% \\ \hline 
        \textbf{PDO Crashes} & -1.67\% & 1.07\% \\ \hline
        \textbf{Injury Crashes} & 5.49\% & -0.46\% \\ \hline
    \end{tabular}
\end{table}

\subsection{Speed Limit Violations and Citations}

\begin{figure}[!h]
    \centering
    \includegraphics[width= 0.48\textwidth, height = 2.3 in]{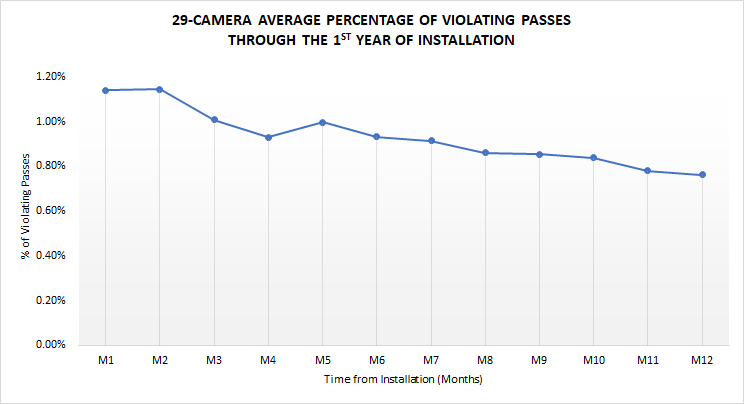}
    \caption{\label{fig:violations} Average percentage of citations over time.}
\end{figure}

Automated speed enforcement systems have always been a subject of public scrutiny due to the fear of unfair citations. To assess this, we analyzed citation data obtained from the Washington D.C MPD to evaluate citation trends for the selected subset of speed enforcement cameras. To account for changing volumes, we looked at citations as a percentage of passing traffic. For the 29 locations of study, the average citation percentage starts at 1.18\% in the first month of installation, steadily decaying to 0.79\% after 12 months as illustrated in Figure \ref{fig:violations}.

\begin{figure}[!h]
    \centering
    \includegraphics[width= 0.48\textwidth, height = 1.7in]{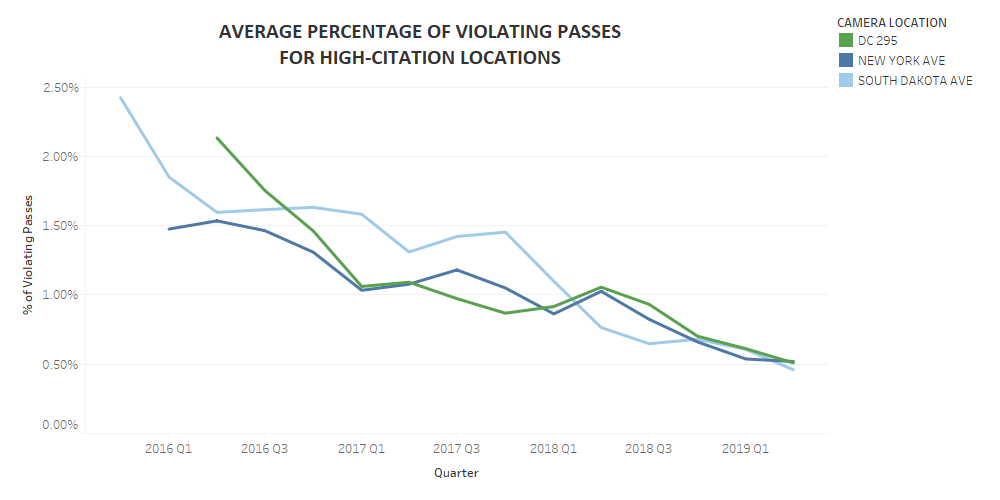}
    \caption{\label{fig:high_violations} Percentage of citations at high-citation locations.}
\end{figure}

We contrast this to three locations with the highest historic citation rates in the District. The three locations were not a part of the 29 sites used for the crash frequency assessment as they have been installed prior to the period of interest in this study. The same trend of decaying citation percentage over time was noticed as shown in Figure \ref{fig:high_violations}, going down to as low as 0.5\% three years after installation.

\begin{figure}[!b]
    \centering
    \includegraphics[width= \textwidth, height= 3.3in]{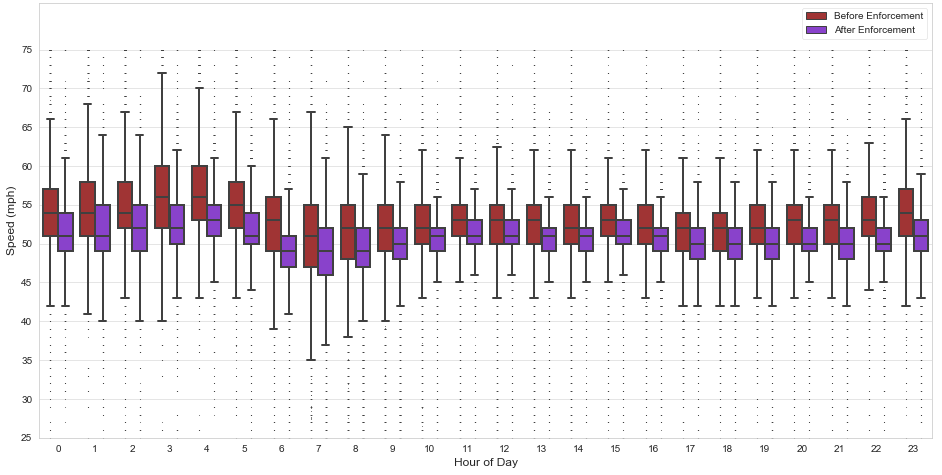}
    \caption{\label{fig:speed} Speed distribution of vehicles in the DC 295 location.}
\end{figure}

\vspace{-13pt}

\subsection{Speeding Behavior}

To further assess the if the decaying citation rate and reduced crash frequency were a result of changes in speeding behavior along the roadway segment and not just at the site of enforcement, we used probe vehicle data to analyze the change in driving speed in the corridor of an ASE camera. The data used contains high granularity speed data (up to one-minute average) across roadway segments. We used the speed data for route which contains the site on DC 295 (also known as the Anacostia Freeway), for which the citations were shown in Figure \ref{fig:high_violations}. Table \ref{tab:295} shows the year-to-year changes (a year prior to installation, and a year after) of speeds in that roadway segment broken down by different percentiles. Figure \ref{fig:speed} shows the hourly distribution of those speeds before and after ASE.

\begin{table}[!h]
  \caption{DC 295 Location Speed Statistics}
  \label{tab:295}
  \centering
  \renewcommand{\arraystretch}{1.8}
   \begin{tabular}{lcc} \hline
        \textbf{Measure} & \textbf{Before Enforcement} & \textbf{After Enforcement} \\ \hline
        \textbf{Mean} & 53.01 & 50.51 \\ \hline
        \textbf{Standard Deviation} & 6.62 & 4.88 \\ \hline
        \textbf{25th Percentile} & 50 & 49 \\ \hline 
        \textbf{50th Percentile} & 53 & 51 \\ \hline
        \textbf{75th Percentile} & 56 & 53 \\ \hline
        \textbf{95th Percentile} & 63 & 57 \\ \hline
        \textbf{99th Percentile} & 72 & 61 \\ \hline
    \end{tabular}
\end{table}

\begin{figure}[!h]
    \centering
    \includegraphics[width= 0.48\textwidth, height = 2.2in]{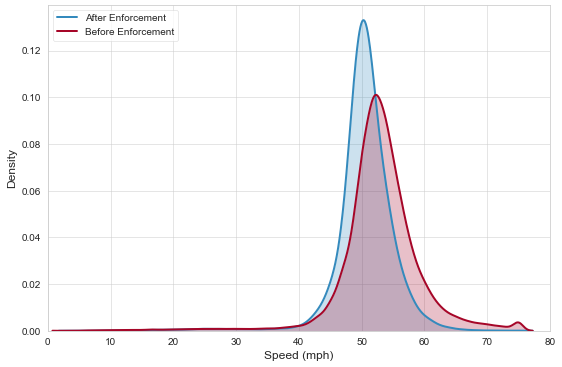}
    \caption{\label{fig:kde} KDE of speed in the vicinity of DC 295 location.}
\end{figure}

ASE cameras in the District issue citations after a passing vehicle exceeds the speed limit by more than 11 mph. The camera on this specific site is enforcing a 50 mph speed limit, hence issuing citations for vehicles whose speed exceeds 61 mph. Looking at the speed data before installation, the top 5\% of the passing vehicles would have been issued citations if they all do not comply. The site had a 2.2\% citation rate in the first month after installation, meaning well over 50\% of those who were previously speeding have complied directly upon enforcement. The speed data also shows that a year after enforcement, 99\% of the drivers were driving at or below the threshold for citations, which conforms with the citation rate falling below 1\% after the first year of installation as obtained from the citation data and illustrated in Figure \ref{fig:high_violations}. A year after enforcement, the average speed on site is equal to the desired and enforced speed limit of 50 mph. The before and after probability distribution of speeds is illustrated in Figure \ref{fig:kde}.

\newpage

\vspace{-9pt}

To reduce errors in speed over-estimation, the speed data provider caps the estimated speeds in the vicinity of the site at 75 mph, hence a bias can be noticed in the probability estimate of 75 mph. The probability distribution of speeds shows the significant shift in speed distribution towards the desired 50 mph limit, accompanied by a significant reduction in speeds above that limit. Similar behavioral shifts were observed only 3 months after installation. Those results provide an insight to practitioners for estimating the citations rate using the existing speed data at a given site before enforcing a speed limit. Gradually decreasing the enforced speed as the behavior change occurs may provide the desired safety benefits while aiding in reducing public skepticism towards ASE.

\section{Discussion and Conclusion}

In this study, we quantified impacts of ASE on crash frequencies, citations, and speeding behavior in select locations within the District of Columbia using traffic crash, ASE citations, and probe vehicle trajectory datasets collected within the District's streets. The results of this exploratory analysis confirm that automated speed enforcement provides an evident contribution to increasing safety and achieving the District's Vision Zero goals. This exploratory analysis used several distinct methods to understand the influence of automated enforcement on driver behavior, in particular on the proclivity of drivers to speed, which has been identified as a critical factor for reducing the number and severity of traffic crashes. By analyzing the pre and post-installation crash patterns near 29 traffic cameras, the study shows a clear reduction in overall reported crashes in comparison to the District's overall trends. In addition, injury crashes – which include the most severe and life-altering crashes on District roads – are reduced even more than less serious crashes.

A key second analysis conducted through this study is the citation rate over time, meaning the percentage of all drivers passing an ASE camera that receive citations. Data from multiple cameras on high-volume roadways shows a clear pattern of decaying citation rates over time, as frequent flyers reduce their speeds to stay within the non-citation range for the ASE program. This matches the pattern identified in the 29 cameras studied for crash analysis, where citation rates decayed from 1.18 percent of all passing drivers to 0.79 percent over the first twelve months after camera activation. Finally, a third analysis evaluates the distribution of speeds at a high-volume ASE site along DC 295 before and after enforcement. While mean speeds declined by less than 3 mph, the most aggressive, fastest drivers in the top 1 percentile reduced their speeds by over 10 mph, from 72 to 61 mph, in response to the presence of the ASE.

The outcomes of this analysis will be used to guide the selection of new installation sites for speed cameras and using the available data in developing performance measures to actively evaluate the impacts of current and future locations. Limitations in data continuity over time, due to the dramatic improvements in crash data reporting in the District since 2016, and the biased traffic patterns in 2020 due to the ongoing pandemic, impinged on the ability of the study to cover multi-year periods before and after installation. Future studies should use the new baseline set through this analysis to review the effect of ASE on crashes over longer periods of time. Identifying and comparing ASE sites to similar sites or corridors with no ASE presence for a more rigorous statistical assessment is also strongly recommended by the authors. Future research should continue to assess the association between traffic speeds at various percentiles and the likelihood of crashes occurring on roadways. Crash analysis methodologies typically require multiple years to lapse before any assessment can be made on an intervention’s safety impact, whereas speed data can be captured and analyzed almost immediately after interventions are conducted given modern data availability.

\section{Acknowledgement}

This study was supported by the Howard University Transportation Research Center. The authors thank HUTRC for their support.

\vspace{-10pt}

\bibliographystyle{ieeetr}
\bibliography{refs}

\end{document}